\documentclass[twocolumn]{webofc}
\usepackage[varg]{txfonts}   
\usepackage{graphicx}
\usepackage{xcolor}
\begin{document}
%
\title{       Effects of interparticle friction on the response of 3D cyclically compressed granular material}

\author{\firstname{Zackery A.} \lastname{Benson}\inst{1}\and
        \firstname{Anton} \lastname{Peshkov}\inst{2}\fnsep\thanks{Corresponding author: apeshkov@ur.rochester.edu} \and
        \firstname{Derek C.} \lastname{Richardson}\inst{3}\and
        \firstname{Wolfgang} \lastname{Losert}\inst{1}
}

\institute{Department of Physics, University of Maryland, College Park, Maryland, USA
\and
           Department of Physics and Astronomy, University of Rochester, Rochester, New York, USA
\and
           Department of Astronomy, University of Maryland, College Park, Maryland, USA
          }

\abstract{%
We numerically study the effect of inter-particle friction coefficient on the response to cyclical pure shear of spherical particles in three dimensions. 
We focus on the rotations and translations of grains and look at the spatial distribution of these displacements as well as their probability distribution functions. We find that with increasing friction, the shear band becomes thinner and more pronounced. At low friction, the amplitude of particle rotations is homogeneously distributed in the system and is therefore mostly independent from both the affine and non-affine particle translations. In contrast, at high friction, the rotations are strongly localized in the shear zone. This work shows the importance of studying the effects of inter-particle friction on the response of granular materials to cyclic forcing, both for a better understanding of how rotations correlate to translations in sheared granular systems, and due to the relevance of cyclic forcing for most real-world applications in planetary science and industry.
}
\maketitle
\section{Introduction}
\label{intro}
The study of dense granular packings is an active field of research in materials science. In these packings, the macroscopic and structural properties rely heavily on the characteristics of the individual grains (friction, stiffness, restitution, etc.). For instance, it is known that friction can change the stability criterion of dense packings of frictional particles for contacts from 6 to 4 \cite{R1} and that the jamming transition is changed from a second-order continuous one to a first-order discontinuous one \cite{R22}. Moreover, previous work has shown that when frictional packings are subject to shear, asymmetric, ordered phases emerge within the material \cite{R2}, a process that is important for understanding granular self-assembly. Additionally, a limit-cycle behavior in cyclically sheared granular packings has been observed for a range of friction values, in which the state of the system remains unchanged upon repeated shear cycles \cite{R3}. 

The majority of these studies place an emphasis on either macroscopic parameters (volume fraction, pressure, etc.)\ or particle positions and displacements \cite{R4,R5,R6,R7,R8,R9,R10}; however, a key quantity that emerges specifically for frictional systems is particle rotations, which increase the number of degrees of freedom of the system to 6 from 3 in three-dimensional systems. Only a few studies have an explicit focus on grain rotations \cite{R11,R12,R13,R14,R15,R16} especially in the case of spherical particles in 3D, where a direct correlation between displacements and rotations is not evident as it is in the case of irregularly shaped particles or two-dimensional systems where the number of constraints to satisfy is lower. However, this does not mean that rotations are not important.  On the contrary: in dense granular assemblies, energy loss is dominated by frictional dissipation that generates rotational motion. At low particle speeds, collisional dissipation is negligible. Understanding rotations is thus essential to comprehend the dynamics of frictional granular systems, even for spherical particles where the relation between translations and rotations is not straightforward.

\begin{figure}[b!]
\centering
\includegraphics[width=1.0\columnwidth]{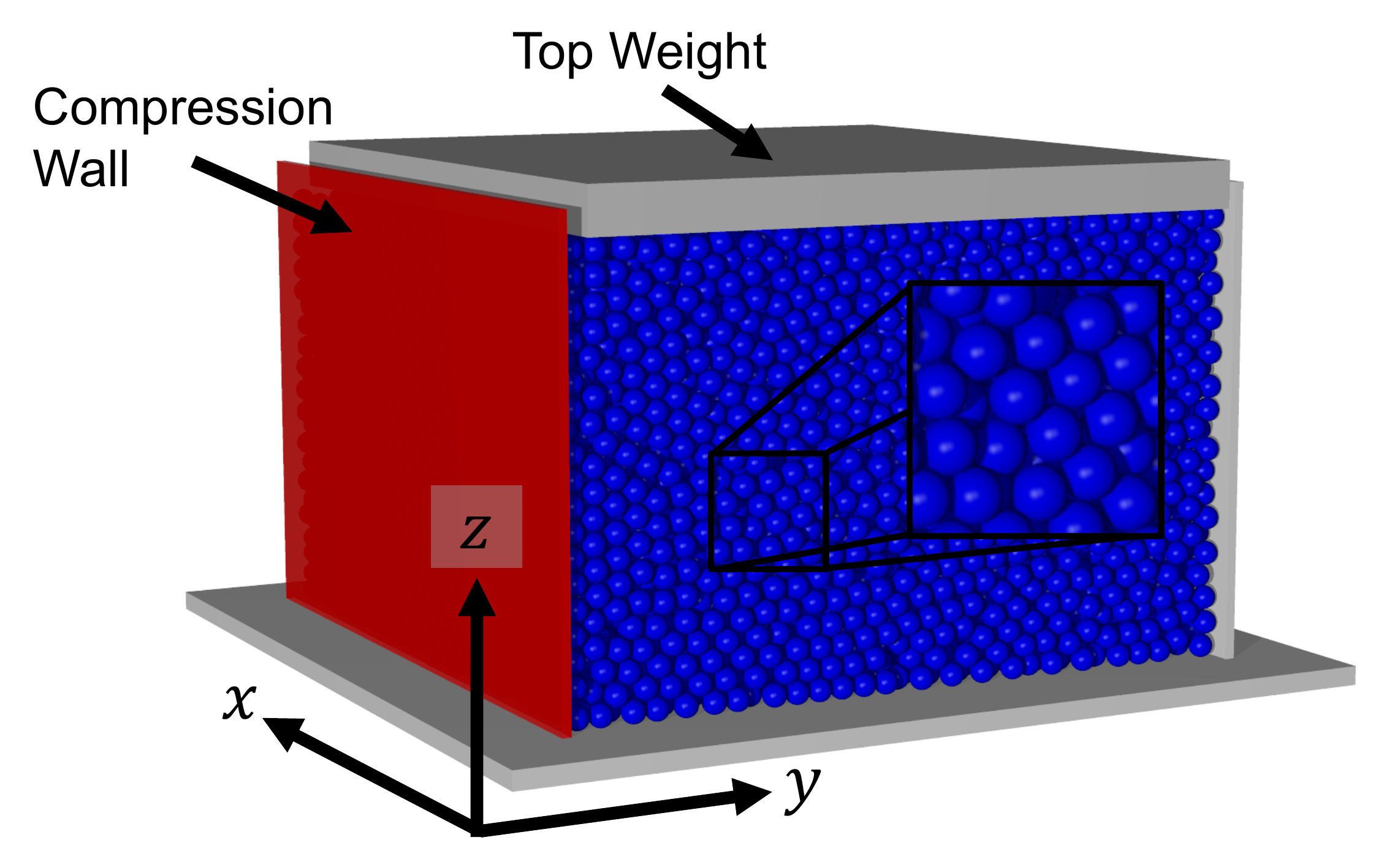}
\caption{Schematic of the simulation setup. 20,000 spheres are confined in a box of side length 15 cm. The red wall compresses the system from 0 (decompressed) to an amplitude $A$ (compressed), along the $y$ axis, then back to 0. A top weight is used for constant pressure. }
\label{schematic} 
\end{figure}

\begin{figure*}[h]
\centering
\includegraphics[width=1.8\columnwidth,clip,trim=0 9.8cm 1cm 10.0cm]{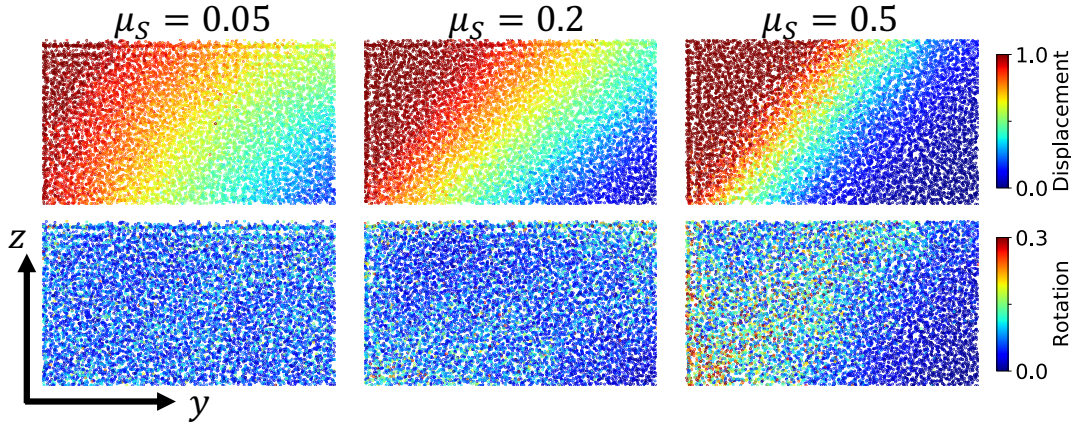}
\caption{Two-dimensional projection of the amplitude of grain translations (top row) and rotations (bottom row) when fully compressed for varying friction coefficients. All values here were normalized by the compression amplitude $A=0.15$~cm. The scale for the rotations is saturated for clarity. The system is compressed from the left. A weight is free to move on the top}
\label{fig-1} 
\end{figure*}

Theoretical studies such as the frictional Cosserat models \cite{R16} have shown the importance of vorticity and rotations in two dimensions to predict the velocity profiles of granular flows. Their predictions have shown a good agreement with experiments and simulations \cite{R18,R23,R24}. However, the model has not yet been linked to three-dimensional particle-scale simulations, and the dependence of the continuum model parameters on the microscopic parameters of the grains, such as interparticle friction, is unknown.

This numerical study is aimed at establishing the connection between translation and rotation of grains in support of related investigations. The first dedicated to the study of how the reversibility of translational motion compares to the reversibility of rotational motion\cite{R12}.  The second investigate how friction and rotations affect the memory formation in granular systems \cite{R13}.

In this paper, we present work focused on quantifying granular rotations of a cyclically compressed 
system using soft-sphere discete-element-method (DEM) simulations, as a function of inter-particle friction between grains. We study how friction affects the localization of shear zones in the system, which in turn affects the spatial distribution of rotations and their correlation to the translations.

\section{Numerical Methods}
Soft-sphere DEM simulations were performed using in-house software called pkdgrav \cite{R19}. 20,000 spherical grains of radius 0.25~cm and mass 0.0618~g were dropped in a cubic box with side length 15.0~cm. A top weight of mass 1000~g was dropped onto the grains to maintain a constant external pressure, and is free to move along the $z$ axis.  The beads settle to a height of approximately 10~cm. A schematic of the simulation setup is given in Figure \ref{schematic}. The system was allowed to settle before starting a quasi-static compression along the $y$ axis. Given the free moving weight on top, the system expands along the $z$ axis when compressed along the $y$ axis, presenting a pure shear geometry. For each cycle, the wall starts off at position 0 (decompressed) then moves to an amplitude $A$ (compressed), then back to position 0. The compression amplitude used in this report is $\sim$1.0\% of the size of the container, corresponding to 0.15~cm. The results presented in this paper are for a system which was subjected to 200 cycles of pure shear.  We focus on small shear amplitudes, where the evolution of the system is small \cite{R12,R13}. The linear displacements are calculated as the difference in position of the grains when the wall is at position 0 to an amplitude $A$. Rotations are defined by a rotation matrix between the orientation of each grain when the wall is at position 0 and compressed to $A$. The units are converted to cm by multiplying the angle of rotation by the radius. We have varied the coefficient of inter-particle friction from $\mu_S$ = 0.05 to $\mu_S$ = 0.5. More details on the numerical setup are available in \cite{R12}. Here we expand on this study and describe key simulation results as a function of friction coefficient.

\section{Results and Discussion}
Figure \ref{fig-1} shows the spatial distribution of displacements and rotations at full compression for 
static friction values of 0.05, 0.2 and 0.5. We see well-defined shear bands in 
the translational displacement for all friction values. The shear band becomes more localized as the friction is increased.
For lower friction, the rotations are uniformly distributed throughout the 
container, behaving more fluid-like as contacts have a lower threshold to slide against one another. 
In contrast, at high friction, the rotations are predominately localized in the region of the translational shear band. This behavior at high friction is reminiscent of that of non-spherical particles \cite{R20}. 

\begin{figure}[ht]
\centering
\includegraphics[width=0.8\columnwidth]{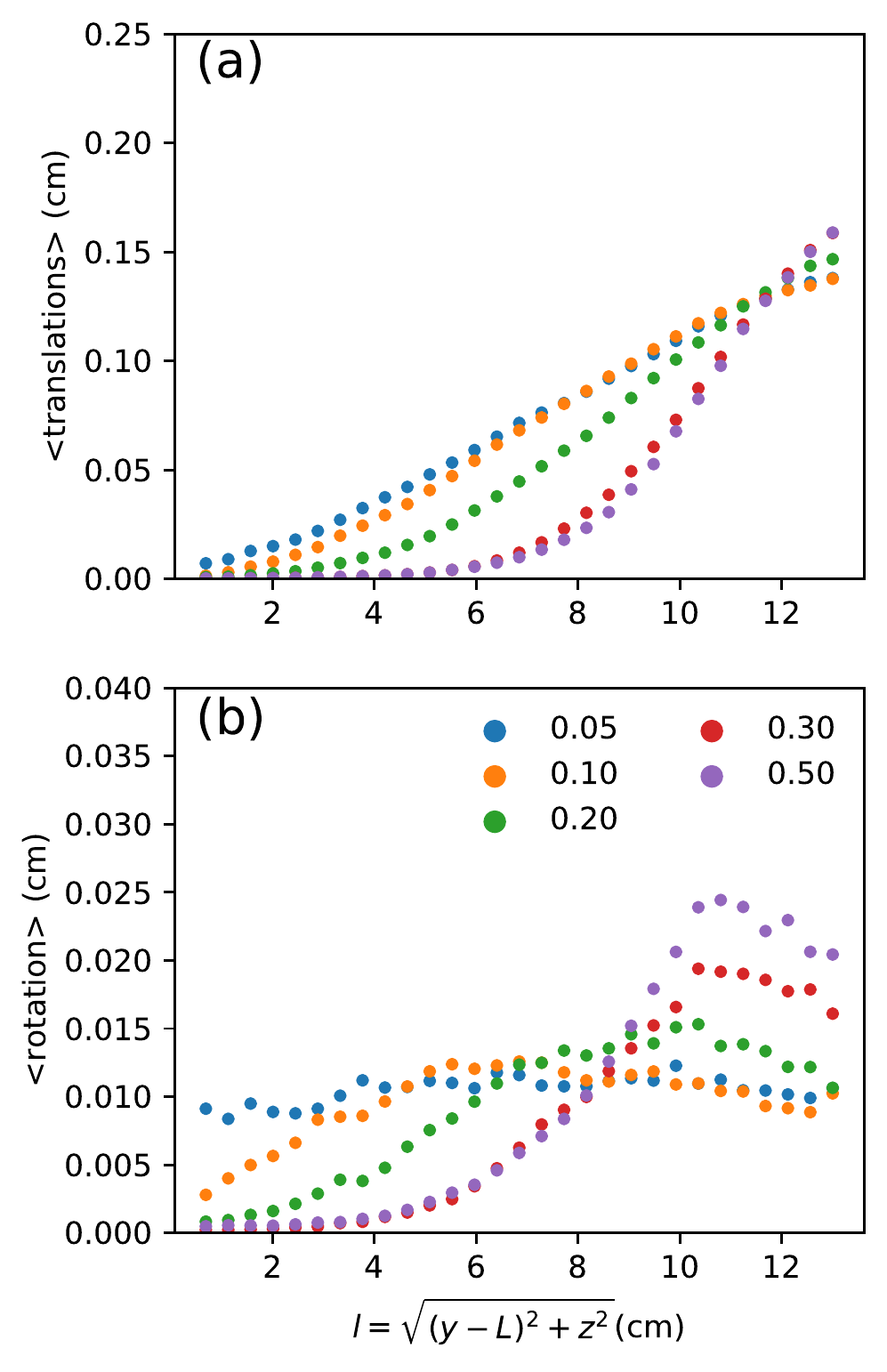}
\caption{Top, the average translational displacement in the binned zones for different values of friction. Bottom, the same for rotations. The binning is performed along an axis parallel to the shear bands. The legend corresponds to the value of friction, $\mu_S$.}
\label{fig-2}  
\end{figure}

\begin{figure}[ht!]
\centering
\includegraphics[width=0.8\columnwidth]{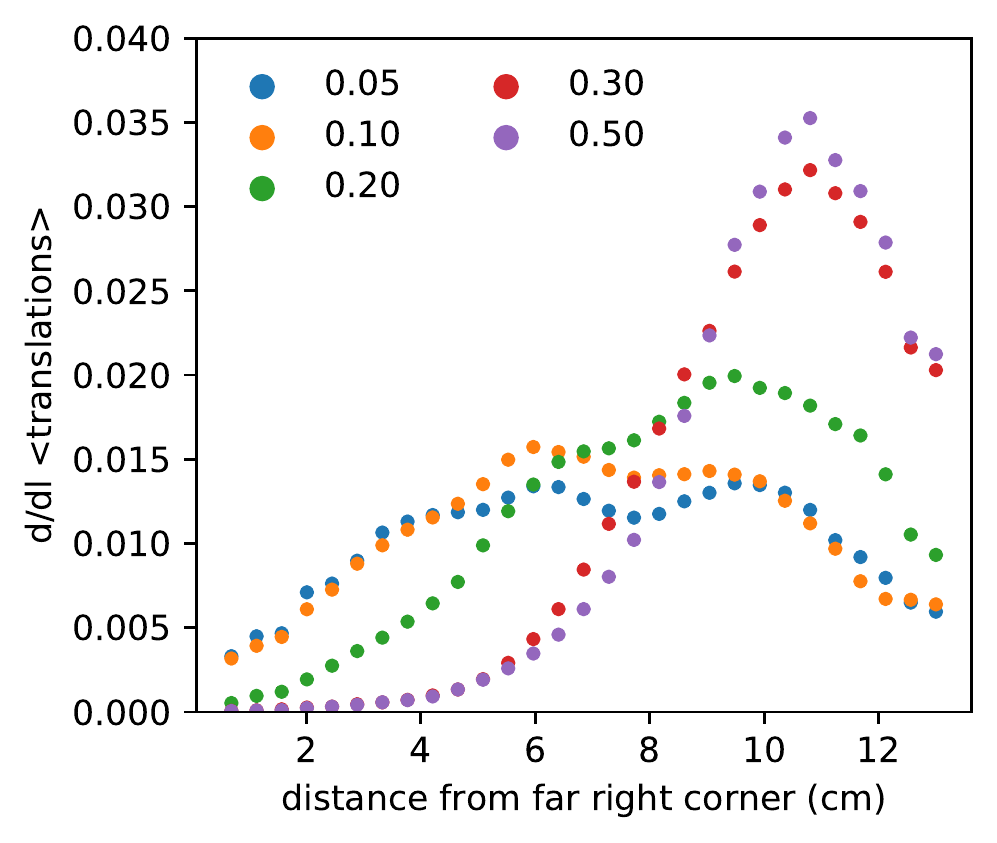}
\caption{Spatial derivative of the translations in the binned zones for different values of friction. The binning is performed along an axis parallel to the shear bands. The legend corresponds to the value of friction, $\mu_S$.}
\label{fig-3}  
\end{figure}

Figure \ref{fig-2} studies the local structure of the shear-zones in Figure 
\ref{fig-1}. Specifically, the displacements were binned along a 45-degree 
line running from the bottom far right of the container ($y,z$ = 0), to the 
top left ($y,z$ = 18~cm) along a line $l=\sqrt{(y-L)^2+z^2}$
; the bins are therefore parallel to the shear band. We crop the extreme 
values of $l$ since the boundaries of the box limits the number of grains 
within those bins. Figure \ref{fig-2}a shows that at low friction, a linear 
increase in the displacements is observed, as expected for a local rheology. 
However, for high friction the curve is highly non-linear, reminiscent of 
what we could expect for non-local rheology \cite{R21}. As we have seen, the 
rotations are uniform at low friction (Figure \ref{fig-2}b). At high 
friction, the rotational displacements are highly localized. We can expect 
that rotations will be the highest in the shear zones, as was observed for 
non-spherical particles \cite{R20}. Accordingly, Figure \ref{fig-3} 
calculates the spatial derivative of the curve in Figure \ref{fig-2}a. 

As for the rotations, we see a peak emerge in the shear zone at high friction. For low values of friction, the function becomes much more uniform, yet is still different from the mostly flat curves of the rotations. This suggests that rotations and translations are mostly decoupled at low values of friction.

\begin{figure}[ht]
\centering
\includegraphics[width=0.9\columnwidth]{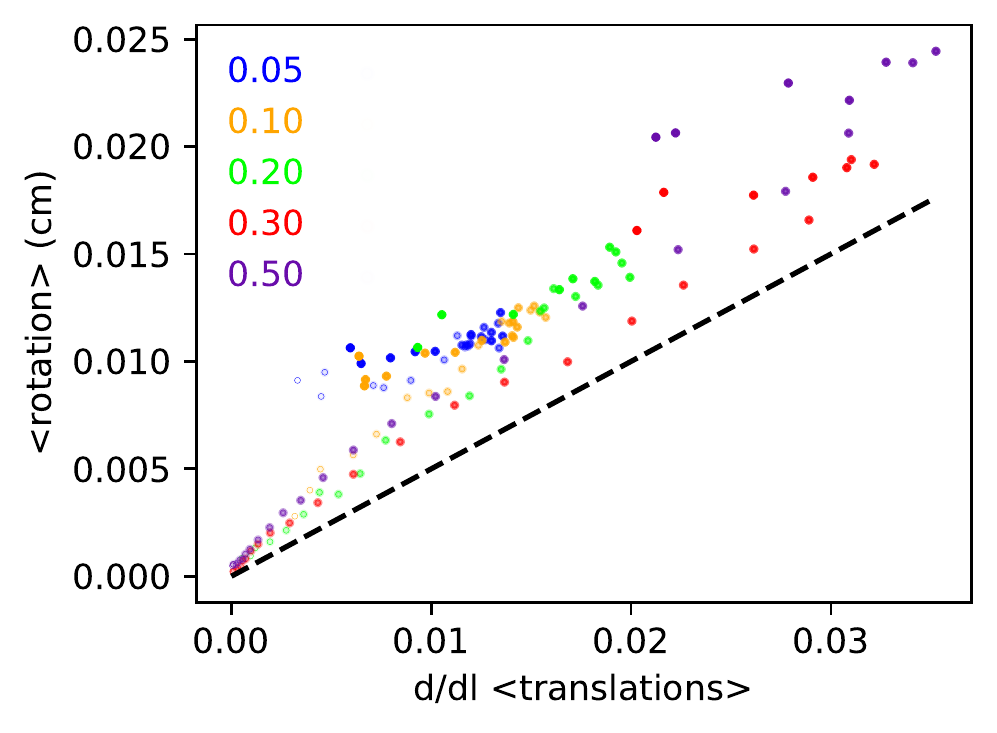}
\caption{Correlation between the spatial derivative of the grain displacements and the rotations for different values of friction. The transparency of symbols corresponds to the distance to the far-right corner as given in Figures \ref{fig-2} and \ref{fig-3}. The dashed line is the theoretical correlation of a sphere rolling between two moving plates, which has a slope of 1/2. The legend corresponds to the value of friction, $\mu_S$.}
\label{fig-4}     
\end{figure}
To further probe the correspondence, the derivative of the displacements and the rotations are plotted against each other in Figure \ref{fig-4}.
For reference, the dashed line corresponds to the relationship between rotation of a sphere rolling without slipping while confined between two parallel plates. In this example, the top plate moves an amount $\Delta x$ in a given time interval and the bottom plate remains fixed; this acts as a measure for the gradient in translations given in Figure \ref{fig-3}. If the sphere cannot slide, then its rotational motion comes out to be exactly $ \Delta x/2$, the relative motion of the two plates. Surprisingly, for high friction, the curve collapses to a straight line of slope $\approx 0.70$. This indicates that there appears to be more rotations than expected for rolling without slipping. Thus, the difference between the dashed line and the data points can be thought of as a measure for frictional dissipation.  In fact, if one takes the sum of the normalized squared distance of each point from the black dashed line, values of 32.5, 24.2, 11.0, 16.8, and 52.2 are obtained with increasing friction. At low friction, the threshold for a contact to slide is much less, which leads to high frictional loss. Increasing the friction also increases the threshold, however the dissipated energy due to sliding is increased. Thus the values that we have computed first decrease and then increase again with friction. 

For the lowest value of friction, $\mu_S = 0.05$, two conclusions can be made. First, rotations do not strongly depend on local shear in this case. Second, there is more rotation than predicted by the amount of local shear, or than observed at higher values of local shear, indicating that rotations are not simply coupled to translational motion. The intermediate friction of $\mu_S = 0.1$ presents an interesting transitional case. While in the region of low translations, the rotations seem to be proportional to the derivative of rotations, as is the case for high friction; in the region of high translations, the two decouple, as is the case for low friction. 

\begin{figure}[t!]
\centering
\includegraphics[width=0.9\columnwidth]{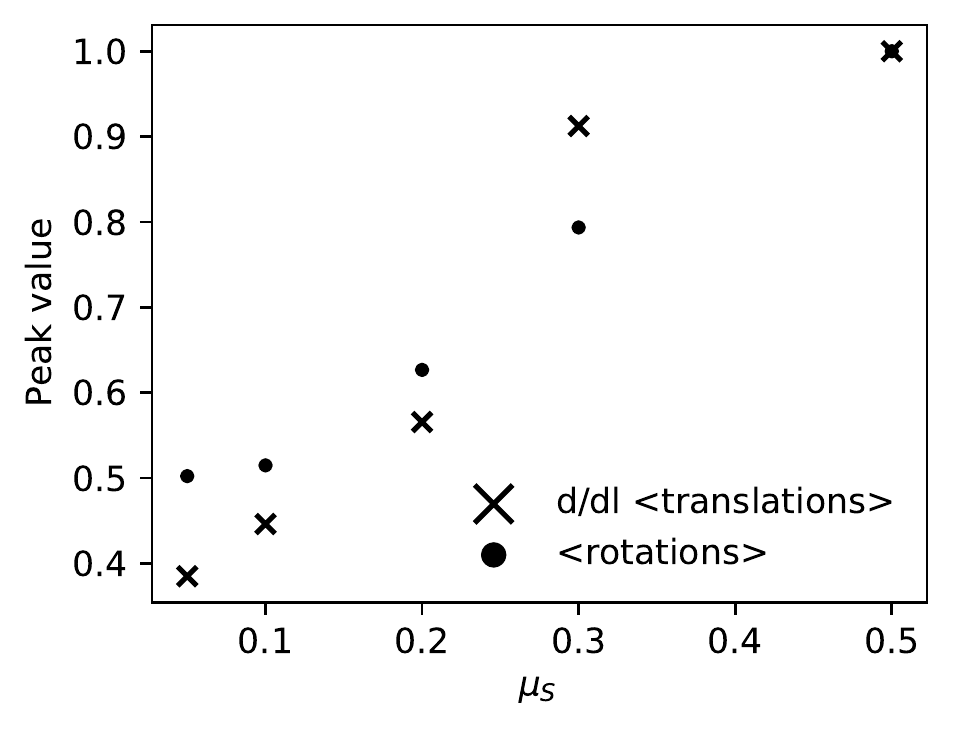}
\caption{Peak value of the derivative of the translations (crosses) and the rotations (circles) as a function of friction. Data points were normalized by the highest peak value for comparison.}
\label{fig-5}
\end{figure}

As the friction increases, the rotations appear to be more confined, as evident in Figure \ref{fig-1} as well as the peaks in Figure \ref{fig-2}. It appears the material goes through some structural transition as a function of friction. To visualize a transition, we plot the maximum value of the rotations curve in \ref{fig-2} as a function of the friction coefficient. Figure \ref{fig-5} shows the value of maximum rotation as well as the maximum of the derivative. The peak values were normalized by the highest quantity ($\mu_S = 0.5$). The evolution seems to be a derivative function with an increase in the peak value between $\mu_S$ = 0.2 and 0.3. As we have already seen on Figure \ref{fig-4}, for low values of friction, there also appears to be more rotations than expected if there would have been a linear proportionality between rotations and the derivative of translations.

\section{Conclusion}
In this article, we have analyzed the influence of inter-particle friction on the response of a three-dimensional granular system of spherical particles to cyclic compression. Three main observations can be made from our results. First, that the shear zone is becoming more localized as the inter-particle friction is increased, a result that agrees with the prediction of frictional Cosserat models \cite{R18}. Second, that at low friction, the rotations are mostly decoupled from translations, while at high friction, the rotations are mostly concentrated in the shear band. The latter result is consistent with that observed for non-spherical particles \cite{R20}. Finally, we find that there is always more rotations than for a theoretical rolling-without-sliding system, which can be used to quantify dissipation in a dense granular system.

\fontsize{9}{12}\selectfont
The authors acknowledge the University of Maryland supercomputing resources (http://hpcc.umd.edu) made available for conducting the research reported in this paper. ZB was supported by the National Science Foundation graduate research fellowship program. WL and AP were 
supported by National Science Foundation Grant No. DMR-1507964. AP was also supported by DMR-1809318.  
%
%
%

\end{document}